\documentclass[manuscript]{aastex}
\usepackage{lineno}
\linenumbers

\shorttitle{Discovery of very high energy gamma-rays from PKS\,1424+240 and multiwavelength constraints on its redshift}
\shortauthors{Acciari et al.}

\begin{document}

\title{Discovery of very high energy gamma rays from PKS\,1424+240 and multiwavelength constraints on its redshift}

\author{
\textbf{VERITAS collaboration:}
V.~A.~Acciari\altaffilmark{v1},
E.~Aliu\altaffilmark{v2},
T.~Arlen\altaffilmark{v3},
T.~Aune\altaffilmark{v4},
M.~Bautista\altaffilmark{v5},
M.~Beilicke\altaffilmark{v6},
W.~Benbow\altaffilmark{v1},
M.~B{\"o}ttcher\altaffilmark{v7},
D.~Boltuch\altaffilmark{v2},
S.~M.~Bradbury\altaffilmark{v8}
% ...
}
\altaffiltext{v1}{Fred Lawrence Whipple Observatory, Harvard-Smithsonian Center for Astrophysics, Amado, AZ 85645, USA}
\altaffiltext{v2}{Department of Physics and Astronomy and the Bartol Research Institute, University of Delaware, Newark, DE 19716, USA}
\altaffiltext{v3}{Department of Physics and Astronomy, University of California, Los Angeles, CA 90095, USA}
\altaffiltext{v4}{Santa Cruz Institute for Particle Physics and Department of Physics, University of California, Santa Cruz, CA 95064, USA}
\altaffiltext{v5}{Physics Department, McGill University, Montreal, QC H3A 2T8, Canada}
\altaffiltext{v6}{Department of Physics, Washington University, St. Louis, MO 63130, USA}
\altaffiltext{v7}{Astrophysical Institute, Department of Physics and Astronomy, Ohio University, Athens, OH 45701}
\altaffiltext{v8}{School of Physics and Astronomy, University of Leeds, Leeds, LS2 9JT, UK}
% ...

\begin{abstract}
We report the first detection of very-high-energy\footnote{$\gamma$-ray emission above 100\,GeV} (VHE) gamma-ray emission above 140\,GeV from PKS\,1424+240, a BL Lac object with an unknown redshift. The photon spectrum above 140\,GeV measured by VERITAS is well described by a power law with a photon index of $3.8 \pm 0.5_\mathrm{stat} \pm 0.3_\mathrm{syst}$ and a flux normalization at 200\,GeV of ($5.1\pm 0.9_\mathrm{stat}\pm 0.5_\mathrm{syst})\times10^{-11}$\,TeV$^{-1}$cm$^{-2}$s$^{-1}$, where $\mathrm{stat}$ and $\mathrm{syst}$ denote the statistical and systematical uncertainty, respectively. The VHE flux is steady over the observation period between MJD 54881 and 55003 (2009 February 19 to June 21). Flux variability is also not observed in contemporaneous high energy observations with the \emph{Fermi} Large Area Telescope (LAT). Contemporaneous X-ray and optical data were also obtained from the \emph{Swift} XRT and MDM observatory, respectively. The broadband spectral energy distribution (SED) is well described by a one-zone synchrotron self-Compton (SSC) model favoring a redshift of less than 0.1. Using the photon index measured with \emph{Fermi} in combination with recent extragalactic background light (EBL) absorption models it can be concluded from the VERITAS data that the redshift of PKS 1424+240 is less than 0.66.


\end{abstract}


\section{Introduction}

PKS\,1424+240 was detected as a radio source by \cite{1977AJ.....82..692C}.
 It was classified as a blazar by \cite{1988ApJ...333..666I} from optical polarization studies. \cite{1993AJ....106.1729F} verified the polarization results and also reported non-thermal X-ray radiation, further strengthening the classification.

Blazar emission is dominated by non-thermal radiation, which is thought to be related to charged particle acceleration near a massive compact object in the center of the host galaxy, or in outflowing relativistic jets. The SED is characterized by two peaks. The lower peak is widely accepted to be synchrotron radiation from relativistic electrons and occurs in the IR to X-ray bands. The higher energy peak is in the gamma-ray band, sometimes at energies as high as a few TeV, and can be created via either inverse-Compton scattering by relativistic electrons or hadronic interactions \citep[for a review see][and references therein]{2007Ap&SS.309...95B}.
The position of the synchrotron peak of PKS 1424+240 has not been measured, but it can be constrained from optical and X-ray data to be between $10^{15}$\,Hz and $10^{17}$\,Hz. Depending on the definition used, PKS 1424+240 is either an intermediate-frequency-peaked BL Lac (IBL) \citep{2006A&A...445..441N} or a high-frequency-peaked BL Lac (HBL) \citep{1996MNRAS.279..526P,FermiLBASSED}.

 Gamma-ray emission from PKS\,1424+240 was not detected by EGRET \citep{1994ApJS...94..551F}, but was recently observed with the \emph{Fermi} LAT pair-conversion telescope \citep{2009ApJS..183...46A,2009ApJ...700..597A}.  The reported flux above 100\,MeV of $(6.2\pm0.8)\times10^{-8}$cm$^{-2}$~s$^{-1}$ and hard spectral index $\Gamma=1.80\pm0.07$ ($dN/dE\propto E^{-\Gamma}$)  triggered VERITAS observations.

The redshift of PKS\,1424+240 is not known. \cite{1995A&A...303..656S} have derived a lower limit on the redshift of $z>$ 0.06 and  \cite{2005ApJ...635..173S} a limit of $z>$ 0.67, both assuming a minimum luminosity of the host galaxy. The latter authors also reported evidence that the ratio of the nucleus to host luminosity is much larger than $100$, which is  typical for BL Lac objects but complicates photometric determination of the redshift.


We report the detection of PKS\,1424+240 in VHE gamma rays and contemporaneous observations with \emph{Fermi}, \emph{Swift}, and the MDM observatory.  Shortly after the VHE discovery \citep{2009ATel.2084....1O}, it was confirmed by the MAGIC collaboration \citep{2009ATel.2098....1T}. This discovery marks the first \emph{Fermi}-motivated  VHE discovery.

\section{Observations and Analysis of VERITAS Data}
The VERITAS observatory, located in southern Arizona at 1.3\,km a.s.l., is described in detail in \cite{weekes} and \cite{holder}.


PKS\,1424+240 was observed with VERITAS between 2009 February 19 and June 21 at zenith angles between $7^\circ$ and $30^\circ$. The observations were performed in wobble mode \citep{1994APh.....2..137F} with a $0.5^\circ$ offset, enabling simultaneous background estimation. About one third of the data were taken during low levels of moonlight.
 About 65\% of the observations were conducted with only three telescopes due to the relocation of one telescope, which began in May and was completed in August 2009. Of the 37.3 hours of data, 28.5 hours survive standard data quality selection.

Events are reconstructed following the procedure in  \cite{2008ApJ...679.1427A}.  The recorded shower images are parameterized by their principal moments, giving an efficient suppression of the far more abundant cosmic-ray background. Two separate sets of cuts are applied to reject background events, hereafter called \emph{soft} and \emph{medium}. These cuts are applied to the parameters \emph{mean scaled width} (MSW), and \emph{mean scaled length} (MSL), apparent altitude of the maximum Cherenkov emission (shower maximum), and $\theta^2$, the squared direction between the position of PKS\,1424+240 and the reconstructed origin of the event. Studies on independent data sets show that a shower-maximum cut significantly improves the low energy sensitivity. \emph{Soft} cuts have a higher sensitivity for sources with soft photon spectra because of a lower energy threshold resulting from a minimum \emph{size} cut of 50 photoelectrons. In the \emph{medium} cuts a minimum \emph{size} cut of 100 photoelectrons is applied. \emph{Size} is a measure of the recorded photoelectrons from a shower and a good indicator of the energy of the primary. For the \emph{soft}-cuts analysis the remaining cuts are MSW $<1.06$, MSL $<1.30$,  shower maximum $>7$\,km, and $\theta^2<(0.14^\circ)^2$, and MSW $<1.04$, MSL $<1.28$, shower maximum $>5$\,km, and $\theta^2<(0.1^\circ)^2$ for the \emph{medium} cuts. The cuts have been optimized \emph{a priori} to yield the highest sensitivity for a source with 5\% of the Crab Nebula gamma-ray flux. The results are independently reproduced with two different analysis packages explained in \cite{2008ICRC....3.1385C} and \cite{2008ICRC....3.1325D}.

\begin{figure}
\epsscale{.80}
\plotone{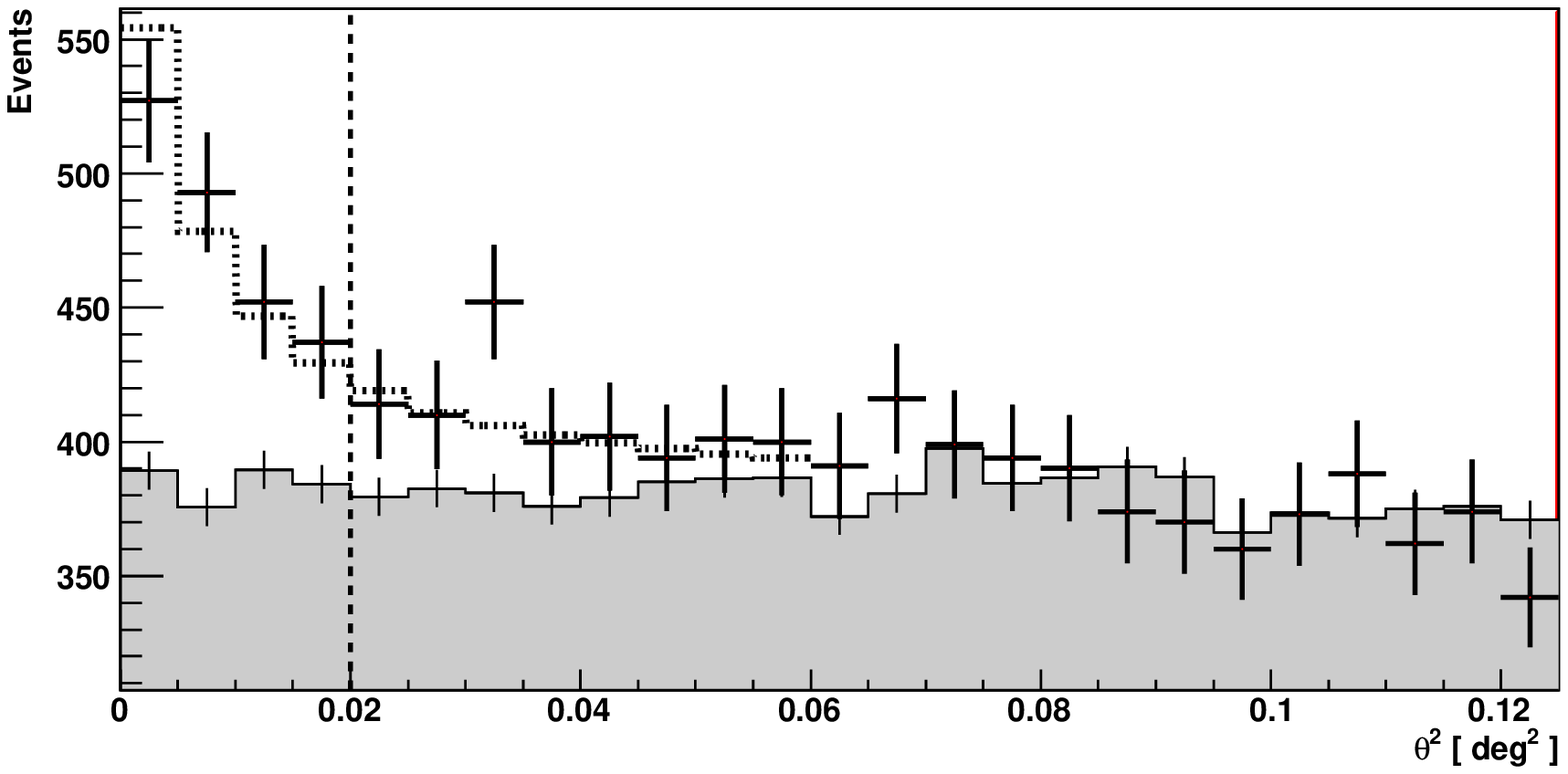}
\caption{\label{thetasqr} Distribution of $\theta^2$ for VERITAS events selected with \emph{soft} cuts. The points with error bars denote the on-source events.  The background is shown by the shaded histogram. The dashed vertical line shows the applied $\theta^2$-cut. The expected distribution for a point source is given by the dotted line.}
\end{figure}

In the \emph{soft}-cuts analysis, 1907 on-source events remain out of $1.25\times10^7$ triggered events. The background calculated with the reflected-region method \citep{2007A&A...466.1219B} is 1537 events, which leaves an excess of 370 events. Figure \ref{thetasqr} shows the corresponding $\theta^2$ distribution. The statistical significance of the observed excess is 8.5 standard deviations, $\sigma$, calculated with Equation 17 of \cite{1983ApJ...272..317L}, and including a trials factor of two for the two sets of cuts. In the \emph{medium}-cuts analysis the post-trials significance is $4.8\,\sigma$  (329 on-source events with an estimated background of 244). The angular distribution of the excess events is consistent with a point source. The center of gravity of the excess is  $14^\mathrm{h}\,27^\mathrm{m},0^\mathrm{s}\pm
 7^\mathrm{s}_\mathrm{stat}$,
 $23^\circ\,47'\,40"\pm2'_\mathrm{stat}$
 coinciding with the position of PKS\,1424+240 in radio \citep{2004AJ....127.3587F}. The VERITAS source name is  VER\,J1427+237.

Figure \ref{lightcurve} shows the light curve of PKS\,1424+240 in different energy bands for the time period overlapping the VERITAS observations. The flux measured by VERITAS above 140\,GeV is $\sim5$\% of the Crab Nebula flux.
 The VERITAS data from each dark period\footnote{The $\sim3$ week observing period between full moons} are combined into a single bin to produce a light curve, which is consistent with a constant flux, $\chi^2=$0.3 for 3 degrees of freedom (d.o.f.). However, even a doubling in flux would have been difficult to detect.  There is no evidence for strong flaring episodes on shorter timescales.

\begin{figure}
\epsscale{.80}
\plotone{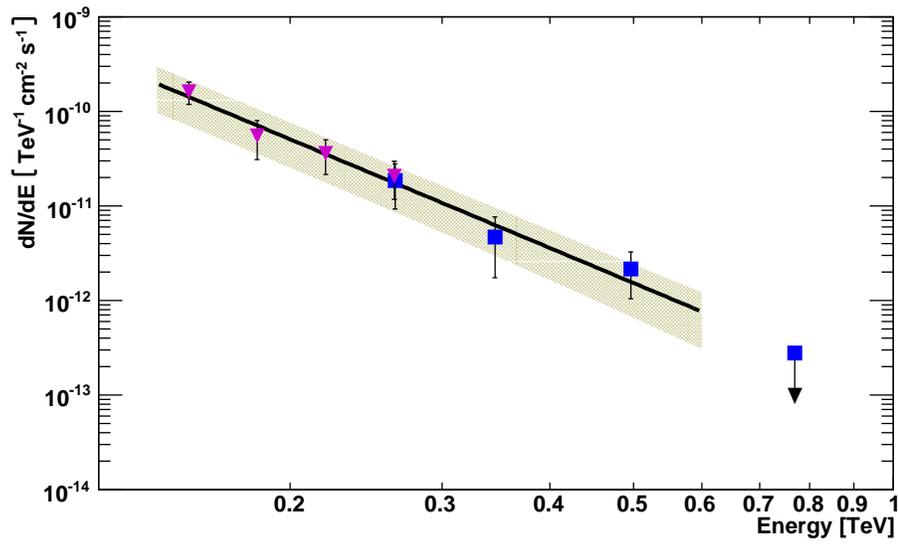}
\caption{\label{spectrum} The time averaged differential photon spectrum of PKS\,1424+240 measured by VERITAS between February 19 and June 21, 2009. The triangles are from the \emph{soft}-cuts analysis and the squares from the \emph{medium}-cuts analysis. The flux point at 260\,GeV is reconstructed in both analysis. The solid lines shows the fit with a power law. The shaded area shows the systematic uncertainty of the fit, which is dominated by a 20\% uncertainty on the energy scale.}
\end{figure}

Figure \ref{spectrum} shows the differential photon spectra derived with the
\emph{soft}-cuts and \emph{medium}-cuts analyses, with one overlapping flux point at 260\,GeV. The fraction of events that are used both in the
last bin in the \emph{soft}-cuts analysis and in the second bin in the
\emph{medium}-cuts analysis is about 2\%, small enough to allow a combined fit
of the flux points from the two analyses, with the more significant \emph{soft}-cuts result at 260\,GeV used in the fit.
The combined spectrum is well parameterized ($\chi^2$=2.2 for 4 d.o.f.) by a power law $dN/dE=F_0 \cdot
\left(E/E_0 \right)^{-\Gamma}$, where the photon
index $\Gamma$ is $3.8 \pm 0.5_\mathrm{stat} \pm 0.3_\mathrm{syst}$ and $F_0$ is ($5.1\pm
0.9_\mathrm{stat}\pm
0.5_\mathrm{syst})\times10^{-11}$\,TeV$^{-1}$cm$^{-2}$s$^{-1}$ for $E_0=200$\,GeV. The combined spectrum is consistent with the fit of the \emph{soft}-cuts points alone, albeit with half the uncertainty on the photon index.

\section{Multiwavelength Observations}


Gamma-ray observations with \emph{Fermi}-LAT (100\,MeV to 300\,GeV), X-ray and optical observations with \emph{Swift} XRT (0.2--10\,keV) and UVOT (170--650\,nm), and optical observations in the R, V and I bands at the MDM observatory  were obtained simultaneously or quasi-simultaneously with the VERITAS observations.

The LAT pair-conversion telescope on board the \emph{Fermi} Gamma-ray Space Telescope continuously monitors the entire sky between 100\,MeV and several hundred GeV \citep{2009ApJ...697.1071A}. The LAT data overlapping with the VERITAS observations were analyzed by selecting ``diffuse'' class events that have the highest probability of being photons. Further event selection was done by only accepting events that come within a 15$^\circ$ radius from PKS\,1424+240 and have energies between 0.1 and 300\,GeV. Events with zenith angles above $105^\circ$ were excluded to limit contamination by gamma rays coming from the Earth's albedo.

The analysis of the photon spectrum and light curve
were performed with the standard likelihood analysis tools
available from HEASARC {\tt ScienceTools v9r15p2}.  Accidental coincidences with charged cosmic rays in the detector were accounted for using instrument response functions {\tt P6\_V3\_DIFFUSE}.  The background model used to extract the gamma-ray signal includes a Galactic diffuse emission component and an isotropic component\footnote{\url{http://fermi.gsfc.nasa.gov/ssc/data/access/lat/BackgroundModels.html}}. The isotropic component includes contributions from the extragalactic diffuse emission as well as from residual charged particle backgrounds.  The spectral shape of the isotropic component was derived from residual high latitude events after the Galactic contribution had been modeled. The background model also takes into account unresolved gamma-ray sources in the region of interest, thus avoiding a bias in the spectral reconstruction. To further reduce systematic uncertainties in the analysis, the normalization and spectral parameters in the background model were allowed to vary freely during the spectral point fitting.

\begin{figure}
\epsscale{.80}
\includegraphics[angle=0, width=18cm]{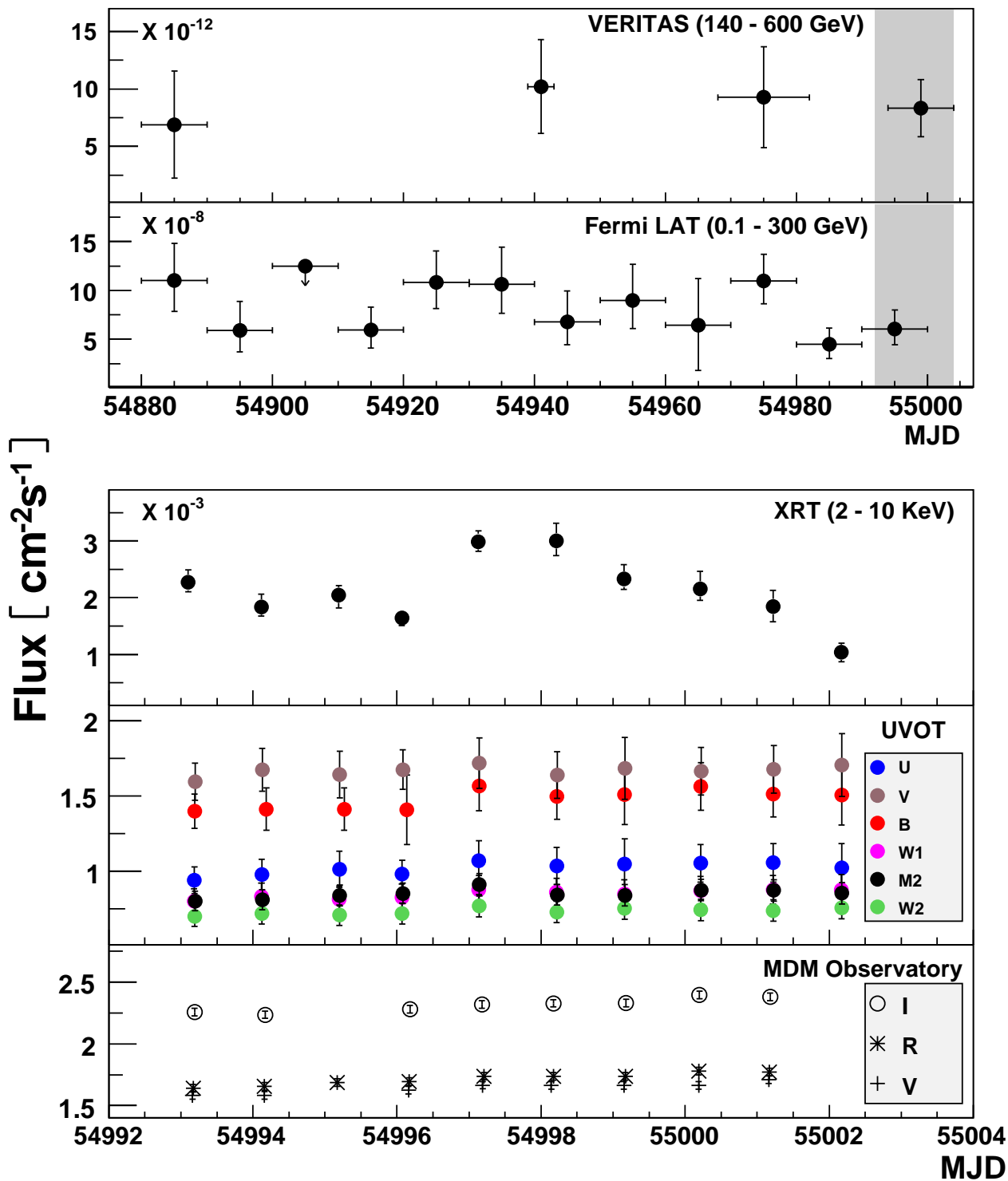}
\caption{\label{lightcurve} Light curves of PKS 1424+240 in VHE gamma rays (VERITAS), HE gamma rays (\emph{Fermi}-LAT), X-rays (\emph{Swift} XRT), UV (\emph{Swift} UVOT) and optical (\emph{Swift} UVOT, MDM). The X-ray, UV and optical light curves cover the time period indicated in the upper two light curves by the shaded region. The horizontal bars in the VHE and HE light curves give the range over which the flux has been integrated. The HE upper limit is at the 95\% confidence level.}
\end{figure}

The \emph{Fermi}-LAT flux measurements are shown in the broadband SED in Figure~\ref{SED}. The flux values are unfolded by assuming an underlying power-law, giving an integrated
flux over the 0.1--300 GeV band
 $(7.04 \pm
0.96_\mathrm{stat}\pm 0.38_\mathrm{sys}) \times 10^{-8}$cm$^{-2}$~s$^{-1}$,
and a differential photon spectral index $\Gamma_{\rm LAT} = 1.73 \pm 0.07_\mathrm{stat}\pm0.05_\mathrm{sys}$.  The light curve of the integral flux above 100\,MeV is plotted with 10-day bins in Figure \ref{lightcurve}. A fit with a constant yields a $\chi^2=11.5$ for 11 d.o.f., suggesting no variability.

Target of opportunity observations of nearly 16\,ksec, distributed over ten observing periods, were obtained with \emph{Swift} \citep{2004ApJ...611.1005G} following the detection of VHE emission from PKS\,1424+240. The data reduction and calibration of the XRT data were completed with the HEASoft v6.6.3 standard tools. The XRT data were taken in photon-counting mode and contained modest pile up for nine of the observations, which was taken into account by masking a region with 3-6 pixel radius around the source. The outer radius chosen for the signal region was 20 pixels and a background region of similar size was chosen about 5 arcminutes off source.

X-ray energy spectra could be extracted from all observing periods and are well described by an absorbed power law using the fixed Galactic column density of neutral hydrogen from \cite{1990ARA&A..28..215D} ($N_\mathrm{h} = 0.264 \times 10^{21}\,$cm$^{-2}$). The fit spectral index varies between 2.1 and 2.9 (photon index between 3.1 and 3.9) with a typical statistical uncertainty of 0.1, while the normalization changes between $1.40\times 10^{-2}$ and $0.74\times 10^{-2}$ photons keV$^{-1}$cm$^{-2}$s$^{-1}$ at 1\,keV with a typical uncertainty of $0.07\times 10^{-2}$ keV$^{-1}$cm$^{-2}$~s$^{-1}$. For the modeling of the SED we use the average spectrum shown in Figure \ref{SED}. The light curve shows that the X-ray flux is variable over the ten days of observation. A fit  to a constant flux yields a $\chi^2$ of 60 for 9 d.o.f..
UVOT observations were taken in the six  V, B U, W1, M2 and W2 bands and were calibrated using standard techniques \citep{2008MNRAS.383..627P}. The reddening has been accounted for by interpolating the absorption values from \cite{1998ApJ...500..525S} with a galactic spectral extinction model \citep{1999PASP..111...63F} obtaining 0.663, 0.968, 0.922 mag for the three UV bands W1, M2, and W2 and an assumed  redshift of z=0. The corresponding light curves are shown in  Figure \ref{lightcurve}.

Data in the optical bands were also obtained with the 1.3\,m telescope and 4K imager of the MDM observatory located on the west ridge of Kitt Peak near Tucson, Arizona. The CCD was operated in unbinned mode, which produces an image scale of 0.315 arcseconds/pixel. 2-4 images were obtained in the V, R, and I bands during each observation. Physical magnitudes were computed from differences in the instrumental magnitudes from the three standard stars in \cite{1996A&AS..116..403F}, assuming that the magnitudes quoted in that paper are exact. The magnitudes were then corrected for Galactic extinction using extinction coefficients calculated following \cite{1998ApJ...500..525S}, taken from NED\footnote{\url{ http://nedwww.ipac.caltech.edu}}, and were then converted into $\nu F_{\nu}$ fluxes. During the 14-day span of the
optical photometry, the visual brightness increased by 14\% and colors became slightly bluer.


\section{Redshift Upper Limit \label{redest}}

 The observed gamma-ray spectrum above 100\,GeV is affected by the absorption of gamma rays via pair conversion with EBL photons \citep{nikishov,1967PhRv..155.1404G}. Depending on the redshift, this effect can result in a significant softening of the spectrum. We estimate an upper limit of the redshift of PKS\,1424+240 by assuming an intrinsic VHE spectrum and making use of the recent advances in extragalactic background light (EBL) modeling.

We assume that the intrinsic spectrum above 140\,GeV can be described by a power law. The hardest photon index that we consider is 1.7, which is the value from the  simultaneous \emph{Fermi} observations. The use of Fermi observations allows a model independent estimate of the hardest possible intrinsic spectrum \citep[see also][]{FermiPG1553}. The power law with an index of 1.7 is absorbed using recent EBL models from \cite{2008A&A...487..837F}, \cite{2009arXiv0905.1144G}, and \cite{2009arXiv0905.1115F}. After absorption the shape of the spectrum is fit to the VERITAS spectrum with the normalization as a free parameter, and the best estimate of the redshift is determined by minimizing $\chi^2$. For an intrinsic index of 1.7 this best fit redshift is z$=0.5\pm0.1_\mathrm{stat}\pm0.1_\mathrm{syst}$ with a $\chi^2$=4  and 5 d.o.f.~. The systematic uncertainty is estimated from the differences in the EBL models.

Instead of assuming no break in the photon spectrum, a more likely scenario is that the intrinsic spectrum softens with increasing energy. In this case an index of 1.7 is an upper limit of the true photon index and the corresponding upper limit on the redshift is $z<0.66$ with a 95\% confidence level.


\section{Spectral Modeling}

\begin{figure}
\epsscale{.80}
\plotone{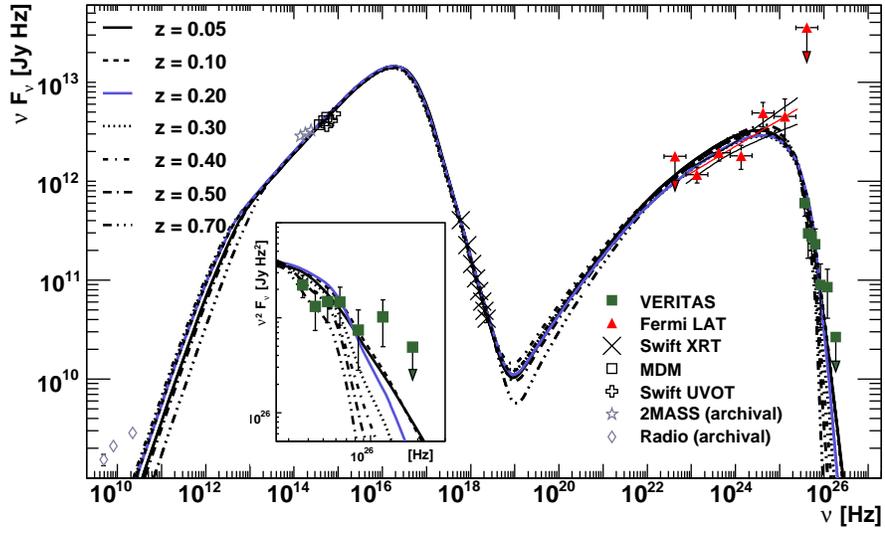}
\caption{\label{SED} SED of PKS\,1424+240. The lines show SSC-model fits assuming different redshifts. The inset shows a zoom of the SED on the VERITAS data in a $\nu^2$F$_\nu$ representation. The \emph{Fermi} data are presented together with their corresponding power-law fit and one standard deviation uncertainty. The upper limits correspond to 95\% confidence level.}
\end{figure}

The spectral energy distribution, comprising data from all of the observations, is shown in Figure \ref{SED}. We model the SED using an improved version of the leptonic one-zone jet model of \cite{2002ApJ...581..127B}. These calculations include time-dependent particle injection and evolution, and they allow for quasi-equilibrium solutions in which a slowly varying broken power-law electron distribution arises from a single power-law injection function, $dn_{\rm inj}/d\gamma \propto \gamma^{-q}$ with a low- and high-energy cutoff  $\gamma_1$ and $\gamma_2$, respectively. All model fits presented here are in the fast-cooling regime, with the cooling break at  $\gamma_1$. We define the magnetic-field equipartition $\epsilon_B$ as $\epsilon_B \equiv L_B/L_e$ with $L_B$ the Poynting flux derived from the magnetic energy density and $L_e$ the energy flux of the electrons propagating along the jet. The corresponding partition fraction for an electron-proton plasma assuming $L_p=10\times L_e$ of cold protons would be one order of magnitude lower.  For an in-depth description of this quasi-equilibrium jet model, see \cite{WComGernot}.

There are few observational constraints on
the model parameters for PKS\,1424+240 and the redshift is unknown. No superluminal motion has been resolved in this object, and it has not been monitored well enough to firmly establish a minimum variability timescale to constrain the size of the emitting region. The different sizes of the emission region $R_B$ assumed here are compatible with the X-ray variability timescale of about a day. We therefore consider a range of plausible redshifts and adopt model parameters which were typically adequate for modeling other VHE blazars. The redshifts investigated range from $z = 0.05$, similar to the redshift of the nearby HBLs Mrk\,421 and Mrk\,501,
to $z=0.7$. This covers the redshift range determined in the previous section and is just above the lower limit set by \cite{2005ApJ...635..173S}, $z > 0.67$.

The shape of the high-energy part of the electron spectrum is well constrained by the rather steep slope of the X-ray spectrum, which has an average photon index $\Gamma_{\mathrm{X-ray}} \sim 3.7$. In all fits, the relativistic electrons are injected into the emission region with a fixed $q = 5.1$.  Lacking direct constraints on the viewing angle $\theta_{\rm obs}$, it was chosen such that the Doppler factor $D = \left( \Gamma [ 1 - \beta_{\Gamma} \cos\theta_{\rm obs}] \right)^{-1} = \Gamma$, where $\Gamma$ is the bulk Lorentz factor of the emitting material, and $\beta_\Gamma c$ is the velocity. The model parameters that were varied are shown in Table \ref{zparameters}. Figure \ref{SED} shows the fits, after EBL absorption using the model of \cite{2009arXiv0905.1144G}.

The SED modeling shows that a reasonable fit can in principle be achieved for any redshift in the considered range. However, the inset in Figure \ref{SED} illustrates that above $z\sim0.2$, the model VHE gamma-ray spectrum becomes increasingly too steep compared with the observed VERITAS spectrum. Furthermore, for redshifts $z>0.4$ the models require unreasonably large Doppler factors of $D > 50$. We note that in particular for the lowest redshift considered, $z=0.05$, a good fit can be achieved with almost equipartition between magnetic-field and electron energy densities.


 An attempt to improve the fit in the gamma-ray bands, by including an external Compton component, results in a steeper VHE gamma-ray spectrum. This is in conflict with the VERITAS spectrum and a worse representation of the \emph{Fermi} spectrum. We therefore conclude that a leptonic fit to the SED of PKS\,1424+240 during the VERITAS observation is possible with a pure SSC model very close to equipartition, in particular if the redshift of the source is $z < 0.1$.

\section{Summary}

We report the detection of PKS\,1424+240 in VHE gamma-rays. The observation with VERITAS was motivated by the release of the first \emph{Fermi} source lists \citep{2009ApJS..183...46A,2009ApJ...700..597A} and this is the first time that \emph{Fermi} observations have led to the discovery of a new source in the adjacent VHE band.

The VHE spectrum of PKS\,1424+240 has a photon index of $3.8\pm 0.5_\mathrm{stat}\pm 0.3_\mathrm{sys}$, whereas the spectrum in the \emph{Fermi} energy range has a photon index of $1.73 \pm 0.07_\mathrm{stat}\pm0.05_\mathrm{sys}$, indicating a break in the spectrum at several tens of GeV.
The break can be explained by a one-zone SSC model assuming a wide range of redshifts or could result from  EBL absorption if the redshift is about 0.5 and the intrinsic photon index is 1.7, from which a redshift upper limit of 0.66 is inferred. The modeling favors a lower redshift but cannot exclude that PKS\,1424+240 is among the most distant sources detected in the VHE regime.


PKS\,1424+240 is the third extragalactic source detected in the VHE regime with an unknown or uncertain redshift. It is evident that increased efforts are needed to determine the redshifts of VHE detected blazars. A redshift measurement will allow a better understanding of the source-intrinsic mechanisms and the absorption effects which go along with the gamma-ray propagation.

\acknowledgments
 VERITAS is supported by grants from the US Department of Energy, the US National Science Foundation, and the Smithsonian Institution, by NSERC in Canada, by Science  Foundation Ireland, and by STFC in the UK. We acknowledge the excellent work of the technical support staff at the FLWO and the collaborating institutions in the construction and operation of the instrument. N.~O.~acknowledges the receipt of a Feodor Lynen fellowship of the Alexander von Humboldt Foundation.

The \textit{Fermi} LAT Collaboration acknowledges support from a number of
agencies and institutes for both development and the operation of the
LAT as well as scientific data analysis. These include NASA and DOE in
the United States, CEA/Irfu and IN2P3/CNRS in France, ASI and INFN in
Italy, MEXT, KEK, and JAXA in Japan, and the K.~A.~Wallenberg
Foundation, the Swedish Research Council and the National Space Board
in Sweden. Additional support from INAF in Italy and CNES in France
for science analysis during the operations phase is also gratefully
acknowledged.

This research has made use of the SIMBAD database, operated at CDS, Strasbourg, France.

{\it Facilities:} \facility{VERITAS}, \facility{Swift}, \facility{Fermi}.

\clearpage

\clearpage

\clearpage

\clearpage

\begin{table}[hbtp]
\caption[]{SSC fit parameters for PKS\,1424+240 as a function of
assumed redshift.}
\label{zparameters}
\begin{center}
\begin{tabular}{ccccccccc}
\hline
{Parameter} & {$z = 0.05$} & {$z = 0.10$} & {$z = 0.2$} & {$z = 0.3$} & {$z = 0.4$} & {$z = 0.5$} & {$z = 0.7$} \\
\hline
$L_e$ [$10^{43}$ erg s$^{-1}$] & 1.60 & 4.12 & 10.7 & 18.9 & 29.2 & 47.1 & 88.8  \\
$L_B$ [$10^{43}$ erg s$^{-1}$] & 1.66 & 5.47 & 16.9 & 31.1 & 45.9 & 49.8 & 66.2 \\
$\gamma_1$ [$10^4$]     & 3.7  & 3.7  & 3.6  & 3.4  & 3.2  & 3.6  & 3.7  \\
$\gamma_2$ [$10^5$]     & 4.0  & 4.0  & 4.0  & 4.0  & 4.5  & 4.0  & 4.0  \\
$D$                     & 15   & 18   & 25   & 30   & 35   & 45   & 60   \\
$B$ [G]                 & 0.37 & 0.31 & 0.25 & 0.24 & 0.25 & 0.18 & 0.14 \\
$\epsilon_B$            & 1.04 & 1.33 & 1.59 & 1.65 & 1.57 & 1.06 & 0.75 \\
$R_B$ [$10^{16}$ cm]    & 1.2  & 2.2  & 3.4  & 4.0  & 4.0  & 4.5  & 5.0  \\
\hline
\end{tabular}
\end{center}
\end{table}

\clearpage

\clearpage

\end{document}